\documentclass[lipsum,onecolumn,amsmath,a4paper,amssymb]{cup-hpl}
\usepackage{graphicx}
\usepackage{siunitx}
\usepackage{color}

\def\Wcm2{W/cm$^2$}

\begin{document}

\shorttitle{Generation of Intense  Deep-Ultraviolet Pulses at 200 nm}
\shortauthor{X. Xie, S. Soultanis, G. Knopp, A. L. Cavalieri, S. L. Johnson}

\title{Generation of Intense Deep-Ultraviolet Pulses at 200 nm}

\author[1]{Xinhua Xie \corresp{xinhua.xie@psi.ch}}
\author[1]{Sharon Soultanis}
\author[1]{Gregor Knopp}
\author[1,2]{Adrian L. Cavalieri}
\author[1,3]{Steven L. Johnson}
\address[1]{Center for Photon Science, Paul Scherrer Institute, Forschungsstrasse 111, 5232 Villigen PSI, Switzerland}
\address[2]{Institute of Applied Physics, University of Bern, 3012 Bern, Switzerland}
\address[3]{Institute for Quantum Electronics, Physics Department, ETH Zurich, CH-8093 Zurich, Switzerland}
\date{\today}

\date{\today}

\begin{abstract}
We report the generation of intense deep ultraviolet pulses at 200 nm with a duration of 48 fs and pulse energy of 130 µJ, achieved via cascaded sum frequency generation using 800 nm femtosecond pulses in barium borate crystals. Efficient frequency up-conversion is realized by optimizing phase-matching conditions and implementing dispersion control, while maintaining the ultrashort pulse characteristics. The generated deep ultraviolet pulses are characterized using two-photon absorption frequency-resolved optical gating, providing detailed insight into their temporal profile and phase. This approach addresses key challenges in ultrashort deep ultraviolet pulse generation, delivering a high-energy, ultrashort source suitable for ultrafast spectroscopy, nonlinear optics, and strong-field physics. These results represent a significant advancement in the generation of high-energy, ultrashort deep ultraviolet pulses, opening new possibilities for time-resolved investigations in ultrafast molecular dynamics, as well as emerging applications in semiconductor science, quantum materials, and photochemistry.
\end{abstract}

\maketitle

\section{Introduction}

The generation of intense and ultrashort deep-ultraviolet (DUV) pulses is essential to advance time-resolved spectroscopy and nonlinear optical studies in the DUV spectral range~\cite{Suzuki2012,Peng2018,Cui2022,ozaki2024}. In particular, sub-100 femtosecond pulses at wavelengths around 200 nm enable the investigation of ultrafast electronic and nuclear dynamics in biomolecules, semiconductors, and novel materials, where strong electronic transitions and resonant nonlinear interactions occur~\cite{Haacke2016,Tsen2004,Fu2025}. The pulse duration is critical in resolving nuclear motion in photon-induced molecular processes, as vibrational wavepackets evolve on femtosecond timescales~\cite{roither2011,xie2014,xie2014_2,erattupuzha2016,larimian2017,xie2015,hu2022}. Shorter pulses provide higher temporal resolution, allowing the direct observation of coherent nuclear wave packet dynamics, conical intersections, and ultrafast photodissociation pathways \cite{BhavyaMuvva2024}. For example, in biomolecular systems such as DNA bases, femtosecond DUV pulses enable the tracking of ultrafast nonradiative relaxation through conical intersections~\cite{Chen2015}. Similarly, in semiconductor nanomaterials, high-energy DUV pulses can be used to probe carrier relaxation dynamics and excitonic effects on ultrafast timescales~\cite{kolesnichenko2024,biswas2022}. Additionally, in strong-field physics, intense sub-100 fs pulses in the DUV regime open new avenues for investigating high-harmonic generation~\cite{Popmintchev2015} and nonlinear optical processes in highly correlated materials~\cite{Saito2015}.

Beyond fundamental spectroscopy, ultrashort DUV pulses have critical applications in seeding high-gain, high-harmonic free-electron lasers (FELs)~\cite{deng2008,Finetti2017,Allaria2015}. The high coherence and short pulse duration of a DUV seed pulse can significantly improve the performance of FELs by reducing temporal jitter and enhancing pulse stability in extreme ultraviolet and soft X-ray regimes~\cite{ackermann2007,Cinquegrana2021}. Seeding FELs with DUV pulses allows for the generation of coherent, high-brightness FEL pulses with improved spectral purity, which is essential for applications in ultrafast atomic, molecular, and condensed matter physics~\cite{mcneil2010}.
The ability to generate intense 200 nm pulses with precise temporal control is therefore a key step in advancing FEL technology and its applications in ultrafast science.

Various approaches have been explored for the generation of femtosecond pulses around 200 nm, each with different trade-offs in terms of energy, duration, and tunability. Frequency up-conversion through harmonic generation and four-wave mixing of femtosecond laser pulses remains a widely used technique, producing pulses in the range from a few to hundreds of femtoseconds~\cite{dubietis2000,durfee1999}. Harmonic generation in gases enables the production of sub-10 fs pulses with extreme spectral broadening in the DUV spectral region, making it highly valuable for time-resolved studies \cite{bauer2005,schultze2008,horio2013}. However, this approach typically yields relatively low pulse energies, limiting its applicability in energy-demanding experiments. Dispersive wave emission in gas-filled hollow-core fibers has demonstrated few-femtosecond DUV pulses at microjoule levels with high repetition rates \cite{kottig2017}. Metasurface-based approaches offer a novel pathway for DUV pulse generation, although they remain in early stages of development~\cite{makarov2016}.

Despite significant progress, achieving high-energy (above 100 $\mu$J) pulses with durations below 100 fs in the DUV remains a technical challenge due to phase-matching constraints, dispersion management, and nonlinear absorption\cite{xuan2018}. In this work, we demonstrate the generation of intense 200 nm pulses with 48 fs duration and 130 $\mu$J pulse energy by frequency upconversion of 800 nm femtosecond pulses using cascaded sum frequency generation in $\beta$-barium borate (BBO) crystals. The pulse energy and pulse duration of fourth-harmonic generation (FHG) is optimized through phase-matching and dispersion control.

\begin{figure}[ht]
\centering
\includegraphics[width=0.8\textwidth,angle=0]{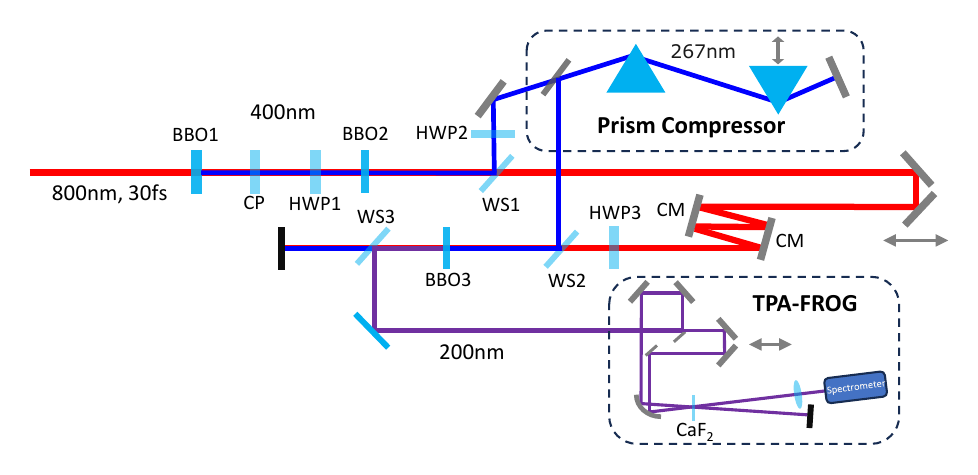}
\caption{Schematic of the experimental setup for fourth-harmonic generation at 200 nm. BBO1: SHG of 800 nm. BBO2: SFG of 800 nm and 400 nm. BBO3: SFG of 800 nm and 267 nm. CP: Time delay compensation plate. HWP1: Half-wave plate for 800 nm (also acts as a full-wave plate for 400 nm). HWP2: Half-wave plate for 267 nm. HWP3: Half-wave plate for 800 nm. WS1, WS2: Wave separators with high reflectivity for 267 nm and high transmission for 400 nm/800 nm. WS3: Wave separator with high reflectivity for 200 nm. CM: chirped mirrors for 800nm.} \label{fig:peda}
\end{figure}

\section{Experiment}

The experimental setup for the generation and characterization of DUV pulses at around \SI{200}{nm} through a three-stage nonlinear frequency conversion process is illustrated in Fig.~\ref{fig:peda} using a collimate input beam. A Ti:sapphire laser amplifier (Coherent Legend Elite Duo HE+) delivers \SI{30}{fs}, 25 mJ pulses at a central wavelength of \SI{800}{nm} with a repetition rate of 100 Hz, which undergo sequential frequency conversion using $\beta$-BBO crystals.

In the first stage, second-harmonic generation (SHG) occurs in the first $\beta$-BBO crystal (BBO1, \SI{200}{\micro m} thickness), where a portion of the \SI{800}{nm} fundamental beam is converted into its second harmonic at \SI{400}{nm}. To compensate for the group velocity mismatch between the generated \SI{400}{nm} and residual \SI{800}{nm} beams, a $\alpha$-BBO compensation plate (CP) is placed immediately after BBO1. A half-wave plate (HWP1) then rotates the polarization direction of the \SI{800}{nm} beam by 90 degrees to ensure Type-I phase matching in the subsequent sum-frequency generation (SFG) stage.

In the second stage, sum-frequency generation takes place in the second $\beta$-BBO crystal (BBO2, \SI{50}{\micro m} thickness with a 2-mm fused silica substrate), where the \SI{800}{nm} and \SI{400}{nm} pulses mix to generate \SI{267}{nm} radiation. A wave separator (WS1) directs the generated \SI{267}{nm} beam while transmitting the residual \SI{400}{nm} and \SI{800}{nm} components. Before further nonlinear conversion, the \SI{267}{nm} beam is sent into a prism compressor, where its group velocity dispersion can be precisely controlled. To minimize reflection losses from the prism surfaces with Brewster incident angle, a half-wave plate (HWP2) is used to change the polarization of the \SI{267}{nm} beam to the vertical direction. Additionally, the polarization of the \SI{800}{nm} beam is rotated by another 90 degrees using a half-wave plate (HWP3), ensuring Type-I phase matching for the final SFG process. A pair of chirped mirrors with four bounces are used for the 800 nm beam to compensate for the group delay dispersion introduced by the optics in the beam path. The 400 nm beam is removed through the 800 nm highly reflective mirrors along the beam path.

In the third and final stage, the \SI{267}{nm} and \SI{800}{nm} pulses are combined in a third $\beta$-BBO crystal (BBO3, \SI{50}{\micro m} thickness with a 2-mm fused silica substrate) via sum-frequency generation, producing the desired \SI{200}{nm} DUV pulses. To avoid the dispersion and two-photon absorption introduced by the fused silica substrate, we place the BBO side in the downstream of the beam path. The time delay between the fundamental and the third harmonic beams is adjusted with a delay stage. A sequence of dielectric wave separators (WS1, WS2, and WS3) selectively transmits or reflects different wavelength components: WS1 and WS2 transmit \SI{400}{nm} and \SI{800}{nm} while reflecting \SI{267}{nm}, whereas WS3 is highly reflective for the generated \SI{200}{nm} pulses. Three additional WS3 elements (not shown) further suppress any remaining undesired harmonic components, ensuring a clean \SI{200}{nm} output.

To characterize the temporal and spectral properties of the generated DUV pulses, a two-photon absorption frequency-resolved optical gating (TPA-FROG) setup is employed. The TPA-FROG technique has been previously demonstrated for visible and infrared pulses \cite{Kim2001TPAFROG, Ogawa2001,Ogawa2002,Ogawa2001a}. FROG is a powerful technique for ultrashort pulse characterization, allowing for accessing both the intensity and phase of a pulse by generating a spectrally resolved nonlinear signal as a function of time delay \cite{Trebino1993,Trebino2000}. In contrast to traditional autocorrelation techniques, which provide only partial temporal information, FROG enables full phase retrieval, making it a highly accurate and widely adopted method for ultrafast optics \cite{Kane1993}. In our experiment, the \SI{200}{nm} pulses are focused into a calcium fluoride (CaF$_2$) crystal with a thickness of 100 $\mu$m, where nonlinear two-photon absorption occurs, generating a signal that varies as a function of the delay between two copies of the pulse. The spectrally resolved signal is then recorded by a spectrometer. Using the FROG retrieval algorithm, the full temporal profile and spectral phase of the pulse can be reconstructed, providing precise measurements of pulse duration \cite{Kane1993}.

\section{Results and Discussion}

\begin{figure*}[ht]
\centering
\includegraphics[width=0.9\textwidth,angle=0]{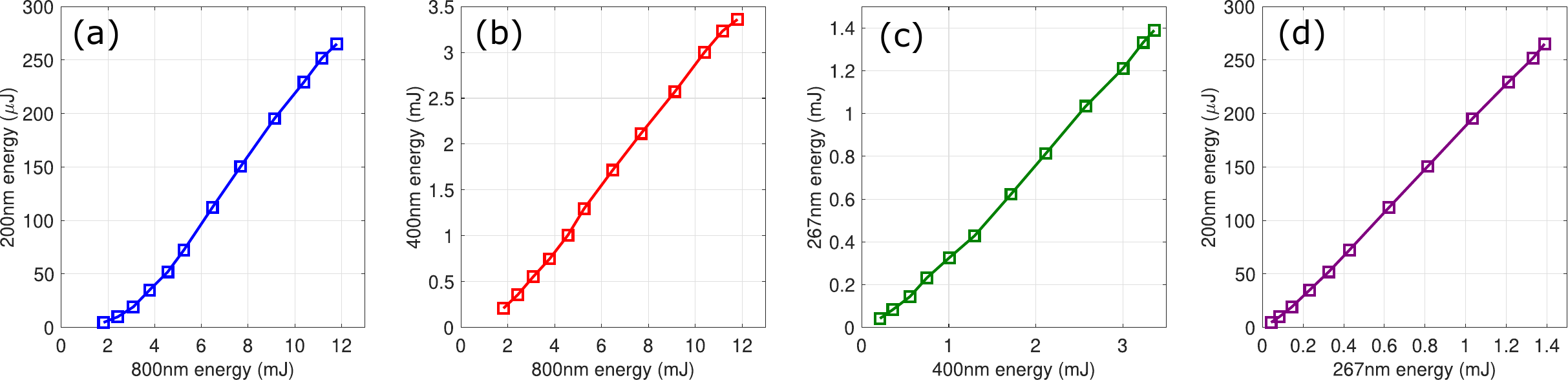}
\caption{Measured pulse energies across the harmonic conversion stages leading to 200 nm generation.
(a) Pulse energy of the 200 nm output as a function of the input 800 nm energy.
(b) Conversion from 800 nm to the intermediate 400 nm second harmonic.
(c) Pulse energy scaling between 400 nm and the generated 267 nm third harmonic.
(d) Final frequency conversion from 267 nm to 200 nm. }
\label{fig:energy}
\end{figure*}

In the experiment, the maximum input pulse energy is set to 18 mJ with a beam diameter of 13 mm. Due to the limited aperture size of the BBO crystals (10 mm in diameter), only 11.8 mJ of pulse energy effectively passes through the nonlinear crystals. As a result, the generated fourth-harmonic pulse at 200 nm achieves a maximal energy of 265 µJ, corresponding to an overall conversion efficiency of 2.2\%.

To characterize the efficiency and energy scaling behavior of the cascaded harmonic generation process, we measured the output pulse energies at each stage of frequency conversion, beginning with the fundamental 800~nm beam and culminating in the final 200~nm output. The results are summarized in Fig.~\ref{fig:energy}, which presents the measured energies across all nonlinear conversion stages. The pulse energy of the 800~nm beam was adjusted by rotating a half-wave plate placed before the 800~nm grating compressor, which effectively acts as a polarizer.

In Fig.~\ref{fig:energy}(a), the 200~nm output energy is plotted as a function of the input 800~nm energy. A nonlinear relationship is observed at low input energies, with the 200~nm energy increasing rapidly at higher fundamental intensities. This behavior is consistent with the high-order nonlinear nature of the FHG process. As the 800~nm pulse energy reaches approximately 5~mJ, the relationship transitions to a linear regime. Saturation effects begin to appear modestly beyond 11 mJ.
Figure~\ref{fig:energy}(b) shows the measured second harmonic (400~nm) energy as a function of the 800~nm input. The energy scaling is nearly quadratic at lower input levels and becomes more linear at higher energies, suggesting the onset of phase-matching saturation or pump depletion effects. The maximum 400~nm output exceeds 3.36~mJ, corresponding to a second harmonic conversion efficiency of over 28\%.
The subsequent THG from 400~nm to 267~nm is presented in Fig.~\ref{fig:energy}(c). In this stage, the 267~nm output scales nearly linearly with the 400~nm input energy, demonstrating stable and efficient THG performance. The maximum output energy reaches approximately 1.39~mJ, with a consistent slope across the entire measured range, indicating favorable phase-matching and low absorption at this wavelength.
Finally, Fig.~\ref{fig:energy}(d) illustrates the energy scaling of the 200~nm output as a function of the 267~nm input energy. A quasi-linear trend is again observed, indicating that the final harmonic conversion step preserves high efficiency with minimal saturation effects. These results validate the sequential harmonic generation scheme and confirm that each stage contributes constructively to the overall system efficiency.
Overall, the cascaded harmonic conversion system demonstrates efficient energy transfer across all stages, with no significant energy bottlenecks. The observed energy scaling behaviors highlight both the high quality of the nonlinear crystals employed and the effectiveness of the temporal synchronization and phase-matching strategies used in the setup.

\begin{figure*}[ht]
\centering
\includegraphics[width=0.9\textwidth,angle=0]{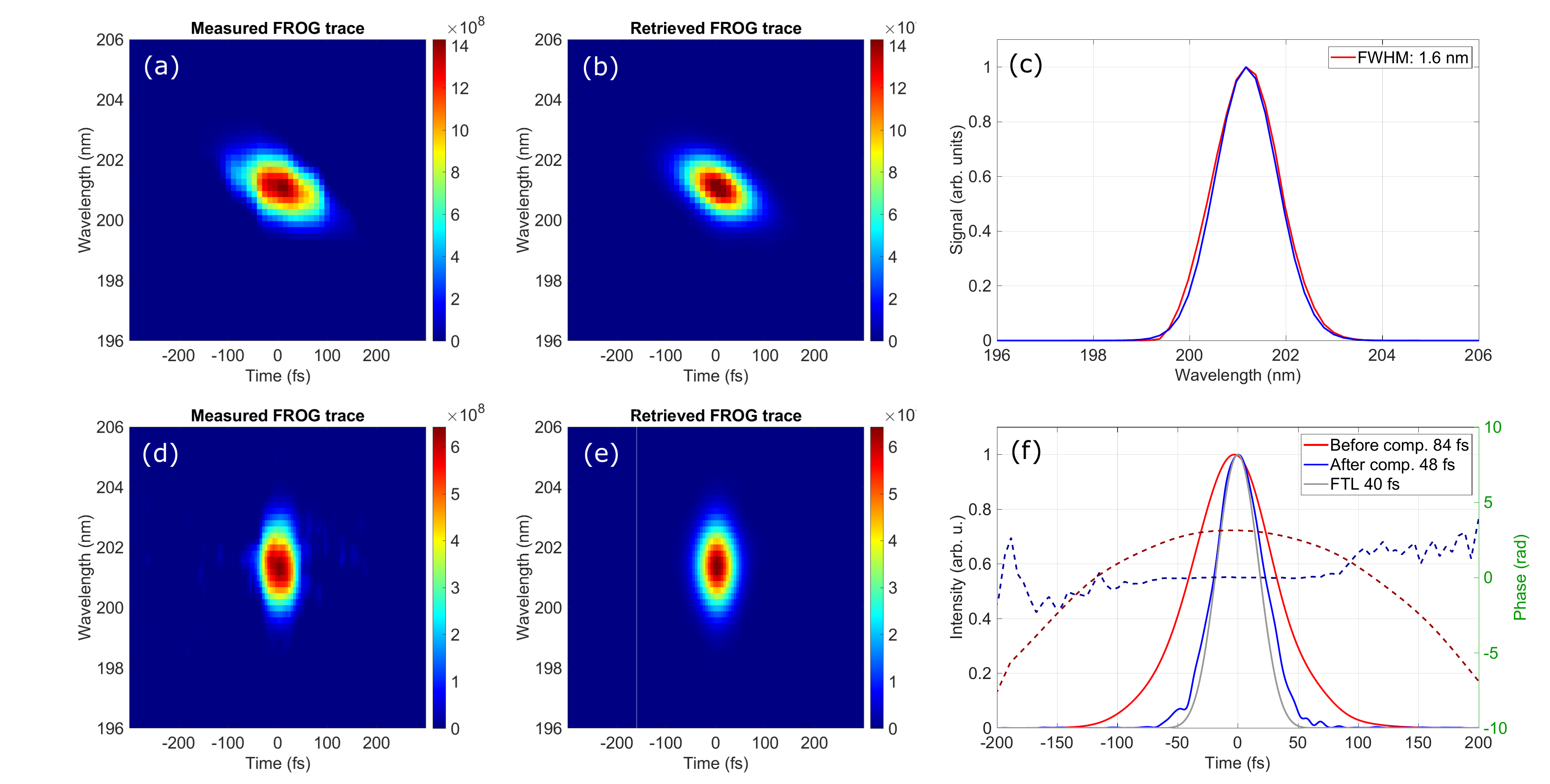}
\caption{The measured and reconstructed SD-FROG traces with the optimization of pulse energy (a,b) and the optimization of pulse duration (d,e) through the dispersion control of the 267 nm beam using the prism pair compressor. (c) Measured and retrieved spectra of the 200 nm beam for the measurement of (a). (f) The reconstructed temporal intensity and phase of the uncompressed and compressed pulses from (a) and (d), together with the temporal profile of the Fourier-transform limited pulse.}
\label{fig:frog}
\end{figure*}

The measured spectrum of the 200 nm pulse, shown in Fig.~\ref{fig:frog}(c), is centered at 201 nm with a full width at half maximum (FWHM) bandwidth of about 1.6 nm, supporting a Fourier-transform limited (FTL) pulse duration of 40 fs. To fully characterize the temporal properties of the pulse, we perform TPA-FROG measurements. The measured and retrieved FROG traces are shown in Fig.~\ref{fig:frog}(a) and (b), respectively. The FROG trace exhibit a noticeable tilt, indicating the presence of significant second-order dispersion in the pulse.
The reconstructed temporal profile and the corresponding spectral phase are presented in Fig.~\ref{fig:frog}(d), yielding a pulse duration of 84 fs. The retrieved parabolic spectral phase indicates a group delay dispersion (GDD) of approximately 1050 fs$^2$. This dispersion originates from the sum frequency generation process in the case of optimizing the 200 nm pulse energy. The presence of residual second-order dispersion highlights the need for further dispersion compensation to achieve even shorter pulse durations, potentially pushing the pulses closer to their transform limit of 40 fs.

Pulse compression of the generated \SI{200}{nm} pulses is investigated using two distinct approaches to optimize the pulse duration while balancing energy loss due to material absorption and nonlinear effects.
The first method employs a CaF$_2$ prism pair to directly compress the 200 nm pulses after generation. Using this setup, we achieve a compressed pulse duration of 49 fs. However, due to the strong two-photon absorption in CaF$_2$ at 200 nm, approximately 92\% of the pulse energy is lost, significantly reducing the overall pulse energy to around 22 µJ.

Given the strong dispersion and absorption exhibited by most optical materials at 200 nm, an alternative approach is explored to achieve self-compression of the 200 nm pulses by controlling the dispersion of the incoming beams. The first strategy involves adjusting the dispersion of the fundamental 800 nm beam using the grating compressor located before the DUV generation setup. This adjustment directly affects the generation of both the 400 nm and 267 nm pulses in the nonlinear crystals. By introducing an additional dispersion of -4760 fs$^2$ to the fundamental beam, we achieved compressed 200 nm pulses with a duration of 46 fs. However, this method results in a pulse energy drop to 28 µJ, mainly due to reduced conversion efficiencies for the generation of the 400 nm and 267 nm with the strongly pre-chirped 800 nm beam.

A more effective approach involves dispersion control of the 267 nm beam alone, using the CaF$_2$ prism compressor prior to the last stage through adjusting the position of the second prism along the beam path, as shown in Fig.~\ref{fig:peda}. This method enables the direct generation of compressed 200 nm pulses from the BBO crystal, resulting in a pulse duration of 48 fs while maintaining a pulse energy of 130 µJ. The overall conversion efficiency from the driving 800 nm to DUV is 1.1\%. The pre-chirp added to the 267nm is about -1520 fs$^2$ according to the position changes of the second prism in the compressor. The measured FROG trace, along with the retrieved trace and reconstructed pulse, are presented in Fig.~\ref{fig:frog} (d, e, f). The vertical shape of the FROG trace indicates minimal residual GDD in the pulse, leading to a flat spectral phase as shown in Fig.~\ref{fig:frog}(e). This flat phase profile confirms effective dispersion management, resulting in near-optimal pulse compression with the pulse duration close to the Fourier-transform-limit.

Importantly, this technique minimizes energy losses while providing better control over the pulse duration, demonstrating a more efficient pathway for generating high-energy, ultrashort DUV pulses. These results highlight the critical role of dispersion management in optimizing pulse compression at 200 nm. The combination of CaF$_2$ prism compression and proper phase-matching conditions achieves a balance between pulse duration and energy, paving the way for enhanced performance in time-resolved spectroscopy, nonlinear optics, and ultrafast molecular dynamics studies in the DUV spectral range.

\section{Beam stability}

\begin{figure}[ht]
\centering
\includegraphics[width=0.6\textwidth,angle=0]{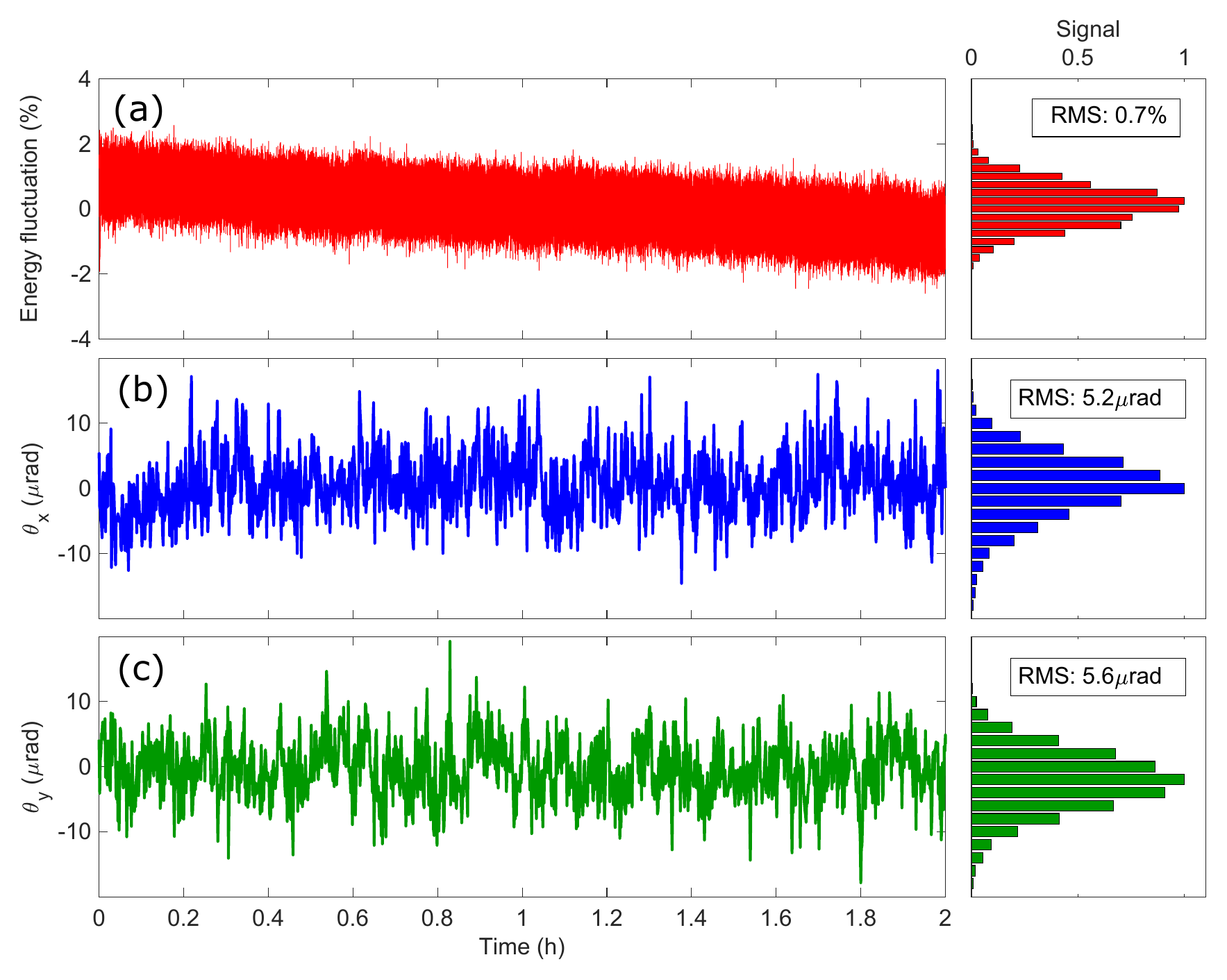}
\caption{Single-shot measurements of pulse energy stability (a), and beam pointing along the horizontal ($\theta_x$) and vertical ($\theta_y$) directions (b), (c) over 2 hours for the compressed 200 nm beam. The histograms of the normalized stability distributions are plotted on the corresponding right-hand side panels.} \label{fig:stab}
\end{figure}

The stability of laser fluence at the interaction point is a critical parameter for ensuring the reproducibility and accuracy of time-resolved measurements. This stability is predominantly governed by the consistency of pulse energy and beam pointing. To maintain optimal experimental conditions over extended operation periods, it is imperative to suppress fluctuations in both parameters.
To assess energy and beam-pointing stability, single-shot measurements of the pulse energy and beam pointing of the compressed fourth harmonic beam were recorded continuously over 2 hours. The Ti:Sapphire laser amplifier exhibited an rms energy stability of 0.15\% over 24 hours for the fundamental beam. The measured pulse energy of the fourth harmonic beam over a 2-hour period, shown in Fig.~\ref{fig:stab}(a), demonstrates an rms stability of 0.7\%, which is well within acceptable limits for most time-resolved applications.
The beam-pointing stability was evaluated by imaging the collimated beam onto a CCD camera positioned at the focus of an uncoated CaF$_2$ lens with a focal length of 100 mm. The beam positions along the horizontal and vertical directions were determined by fitting the recorded beam profile to a 2D Gaussian function. Angular pointing deviations were then calculated by dividing the variations in peak position by the focal length of the lens. The resulting angular pointing stabilities of the compressed DUV pulses were 5.2 $\mu$rad (rms) horizontally and 5.6 $\mu$rad (rms) vertically, as shown in Fig.~\ref{fig:stab}(b, c).
These results confirm that the DUV beam exhibits excellent stability in both energy and beam pointing, making it well-suited for demanding ultrafast spectroscopy and nonlinear optical applications.

\section{Comparison}

\begin{figure}[ht]
\centering
\includegraphics[width=0.6\textwidth,angle=0]{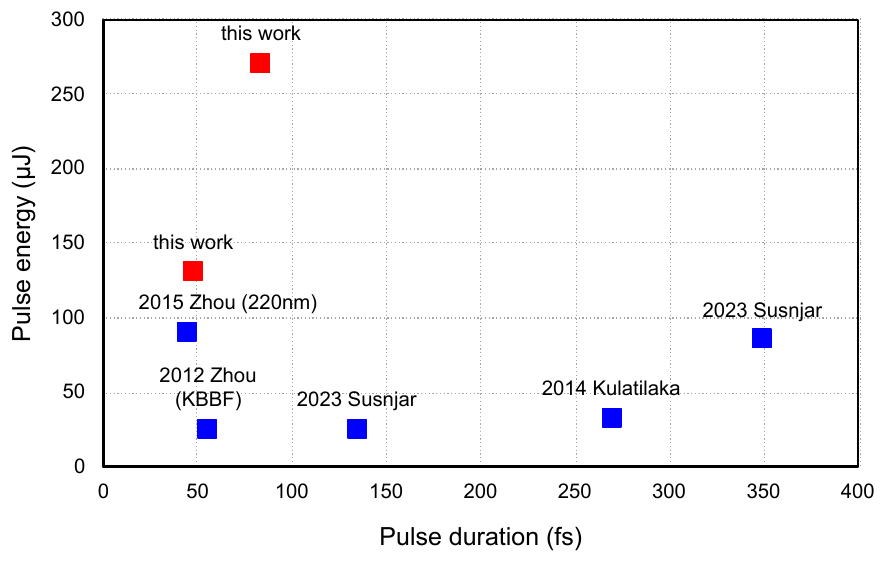}
\caption{A summary of pulse durations and energies for experimentally demonstrated intense 200 nm pulses with all-solid setups.} \label{fig:comp}
\end{figure}

Figure~\ref{fig:comp} compares the pulse duration and pulse energy of intense DUV pulses near 200 nm achieved in this work using cascaded SFG with previous results from all-solid-state platforms \cite{Zhou2015,Zhou2012,Kulatilaka2014,Susnjar2023}. The referenced studies include KBBF-crystal-based generation \cite{Zhou2012}, dual broadband frequency doubling (BFD) \cite{Zhou2015}, cascaded SFG \cite{Kulatilaka2014}, and a common-path fourth-harmonic generation approach \cite{Susnjar2023} using BBO crystals.
Notably, for a wavelength near 220 nm, Zhou et al. achieved a pulse duration of 45 fs and a conversion efficiency of 1.5\% with the dual BFD technique. However, the DUV gratings required for this method may limit its applicability to wavelengths down to 200 nm. In contrast, our approach pushes the energy frontier of DUV pulses to the 100 µJ level while maintaining sub-100 fs durations. By utilizing precise dispersion management of the intermediate 267 nm beam, we further compress the pulses to sub-50 fs durations. This results in enhanced temporal resolution for ultrafast spectroscopy and nonlinear optical experiments.
These results highlight the efficiency, stability, and scalability of our method in generating intense, ultrashort DUV pulses, marking a significant advancement over existing techniques.

\section{Conclusion and outlook}

We have demonstrated a compact, all-solid-state platform for the generation of intense, sub-50fs DUV pulses at 200 nm with pulse energies up to 130 $\mu$J, achieved via cascaded sum-frequency generation in BBO crystals pumped by 800 nm femtosecond pulses. By optimizing phase-matching conditions, implementing precise dispersion control, and minimizing nonlinear absorption, the system achieves high conversion efficiency while preserving pulse duration and beam quality. The resulting DUV pulses are well-suited for pump–probe spectroscopy, strong-field physics, and nonlinear optical studies in the DUV regime.

Comprehensive temporal and spectral characterization using TPA-FROG proved critical, particularly given the lack of conventional nonlinear characterization techniques in the DUV range. TPA-FROG provided accurate pulse duration measurements and exposed residual phase distortions, guiding iterative refinement of the dispersion compensation scheme.

This work addresses key limitations of existing DUV generation schemes by providing a well-balanced approach that combines high pulse energy, ultrashort duration, and system simplicity. Compared to high-order harmonic generation, dispersive wave emission in photonic crystal fibers, and gas-based filamentation, our cascaded-SFG technique offers improved energy conversion and effective pulse compression, while maintaining straightforward alignment and operational robustness. Its inherent scalability and stability present promising opportunities for time-resolved studies in molecular spectroscopy, nonlinear optics, and strong-field physics across the DUV spectral range.

Future efforts will aim to advance pulse compression strategies, broaden wavelength tunability, and embed this source into experimental platforms for next-generation ultrafast spectroscopy~\cite{xie2021,xie2023,xie2024}. Moreover, prospects for scaling pulse energy and repetition rates will be explored to meet the growing demands of high-throughput experimental workflows. Together, these developments represent a significant step toward the realization of high-energy, sub-50 fs DUV sources, poised to unlock new frontiers in ultrafast science across disciplines.

\bibliography{duv}

\bibliographystyle{unsrt}

\end{document}